\documentclass[journal=jctcce,manuscript=article,layout=traditiona]{achemso}

\usepackage[T1]{fontenc}
\usepackage[english]{babel}
\usepackage{subfiles}
\usepackage{amssymb}
\usepackage{amsmath}
\usepackage{amsfonts}
\usepackage{todonotes}
\newcommand{\ve}{\boldsymbol}

\usepackage[version=3]{mhchem}

\title{InfleCS: Clustering Free Energy Landscapes with Gaussian Mixtures}
\author{Annie M. Westerlund}
\affiliation[KTH]{Science for Life Laboratory, Department of Applied Physics, KTH Royal Institute of Technology, Box 1031, SE-171 21 Solna}

\author{Lucie Delemotte}
\affiliation[KTH]{Science for Life Laboratory, Department of Applied Physics, KTH Royal Institute of Technology, Box 1031, SE-171 21 Solna}
\email{lucie.delemotte@scilifelab.se}

\abbreviations{InfleCS,FE,GMM,MD}

\date{\today}
\begin{document}

\maketitle
\begin{abstract}
Free energy landscapes provide insights into conformational ensembles of biomolecules. In order to analyze these landscapes and elucidate mechanisms underlying conformational changes, there is a need to extract  metastable states with limited noise. This has remained a formidable task, despite a plethora of existing clustering methods.

We present InfleCS, a novel method for extracting well-defined core states from free energy landscapes. The method is based on a Gaussian mixture free energy estimator and exploits the shape of the estimated density landscape. The core states that naturally arise from the clustering allow for detailed characterization of the conformational ensemble. The clustering quality is evaluated on three toy models with different properties, where the method is shown to consistently outperform other conventional and state-of-the-art clustering methods. Finally, the method is applied to a temperature enhanced molecular dynamics simulation of Ca\textsuperscript{2+}-bound Calmodulin. Through the free energy landscape, we discover a pathway between a canonical and a compact state, revealing conformational changes driven by electrostatic interactions.
\end{abstract}

\section{Introduction}
Gaussian mixture models provide accurate estimates of free energy landscapes~\cite{westerlund_inference_2018}. Determining metastable core states within a protein's free energy landscape is key to obtaining important biological insights. However, extracting such states from molecular dynamics (MD) simulations with conventional clustering methods is far from straightforward. 

First of all, we are interested in the metastable configurations at free energy minima, the so-called core states. Since proteins move continuously as they explore free energy landscapes, it is difficult to assess an exact state boundary. Moreover, configurations on transition pathways between metastable states generally contribute to noise when characterizing these states. On top of this, the original data is high dimensional, and the necessary dimensionality reduction results in poorly separated states. Finally, the number of metastable core states is typically not known a priori. Thus, to robustly characterize states without any knowledge of the conformational ensemble, we need a clustering method that is solely based on the data. 

Many popular clustering methods are based on simple geometric criteria~\cite{sorensen_method_1948, sneath_application_1957,ward_hierarchical_1963,macqueen_methods_1967}. K-means and agglomerative-Ward, for example, attempt to minimize the within-cluster variance. They work very well on datasets with well-separated spherical clusters, but fail when these assumptions are not met. Spectral clustering~\cite{ng_spectral_2001}, on the other hand, can accurately assign labels to nonconvex clusters by performing spectral embedding prior to K-means clustering. The spectral embedding involves learning the data manifold using local neighborhoods around data points.

In general, geometric clustering methods assign labels to all points and may not accurately identify the boundary between states at the free energy barrier, which leads to noisy state definitions. An idea is to use the data density to identify clusters and make cuts at free energy barriers, as done by for example Hierarchical DBSCAN~\cite{campello_density-based_2013}, density peaks advanced clustering~\cite{rodriguez_clustering_2014,derrico_automatic_2018}, and robust density clustering~\cite{sittel_robust_2016}. Although the idea may appear simple, there are still problems to address. The choice of density basis functions, for example, greatly affects density estimation; local discrete basis functions tend to overfit the density and therefore yield poor estimates in sparsely sampled regions~\cite{westerlund_inference_2018}. Part of the introduced error can be decreased by either using adaptive basis functions~\cite{rodriguez_computing_2018}, or by optimizing the model on the full dataset, as done by Gaussian mixture models (GMMs)~\cite{dempster_EM_1977,bishop_pattern_GMM_2011}. A Gaussian mixture model, however, rests on the assumption of Gaussian shaped clusters. The number of Gaussian components is usually chosen based on how well the model fits the density, which does not necessarily coincide with the number of clusters. Therefore, methods for merging components in GMM to find the correct number of clusters have been proposed~\cite{hennig_merging_2010}.  Another problem is the definition of core state boundaries, which typically are determined with a chosen cutoff~\cite{ester_density_based_1996,rodriguez_clustering_2014}. Such a cutoff does, however, not account for the possibly varying structure and hierarchical nature of a protein free energy landscape.

In this paper, we propose a clustering method that makes minimal assumptions about cluster shapes or dataset structure. We call it the inflection core state (InfleCS) clustering. The functional form of the density landscapes estimated from Gaussian mixtures models is exploited to extract well-defined core states. We show that InfleCS outperforms conventional methods on three different types of toy models, describes its properties with various challenging datasets, and use it to characterize the conformational landscape spanned by molecular simulations of Ca\textsuperscript{2+}-bound Calmodulin. 
\section{Clustering Gaussian mixture free energy landscapes}
The density landscape obtained from a Gaussian mixture model (GMM) estimator~\cite{dempster_EM_1977,westerlund_inference_2018} is used to extract core states with InfleCS. These core states correspond to metastable states in a free energy landscape along collective variables (CVs), $x\in \mathbb{R}^{N_\text{dims}}$,
\begin{equation}
G(x) = -k_BT\log \rho_{\ve{a},\ve{\mu},\ve{\Sigma}}(x),
\label{eq:free_energy}
\end{equation}
where $\rho_{\ve{a},\ve{\mu},\ve{\Sigma}}(x)$ is the Gaussian mixture density at $x$.

\subsection{Gaussian mixture model density estimation}
A Gaussian mixture density is described by a sum of Gaussians with individual amplitudes, $\ve{a}:= (a_i)_{i=1}^{N_\text{basis}}$, means $\ve{\mu}:=(\mu_i)_{i=1}^{N_\text{basis}}$ and covariances $\ve{\Sigma}:=(\Sigma_i)_{i=1}^{N_\text{basis}}$,
\begin{equation}
\rho_{\ve{a},\ve{\mu},\ve{\Sigma}}(x) = \sum\limits_{i=1}^{N_{\text{basis}}} a_i \; \mathcal{N}(x|\; \mu_i,\; \Sigma_i),
\label{eq:GMM_density}
\end{equation}
where $\mathcal{N}(x|\; \mu_i,\; \Sigma_i) = \frac{1}{\sqrt{(2\pi)^{N_\text{dims}} |\Sigma_i |}} e^{\hat{f}_i}$ is a Gaussian with inner function $\hat{f}_i = -\frac{(x-\mu_i)^T \Sigma_i^{-1}(x-\mu_i)}{2}$. 

The parameters of Gaussian mixture models are optimized iteratively with expectation-maximization~\cite{dempster_EM_1977,scikit-learn}. The log-likelihood, $\mathcal{L}$, of the trained data will increase with increased number of parameters. At some point, however, adding more parameters will lead to overfitting. Contrary to what was done in our original GMM free energy estimator~\cite{westerlund_inference_2018}, the number of Gaussian components, $N_\text{basis}$, that allows for a detailed description of the density without overfitting, is selected using the Bayesian information criterion (BIC),~\cite{mclaclan_number_2014, schwarz_BIC_1978}. BIC adds a penalty to the log-likelihood that grows with model complexity,
\begin{equation}
I_{BIC} = \log(N_{\text{points}}) N_{\text{param}} - 2\mathcal{L}.
\end{equation}
To select a final model, we fit multiple GMMs to the data for a given range of $N_\text{basis}$ values and evaluate the $I_{BIC}$ corresponding to each GMM. The model that gives rise to the smallest $I_{BIC}$ is ultimately chosen as the one with best fit.

\begin{figure}
\centering
\includegraphics[width=0.6\paperwidth]{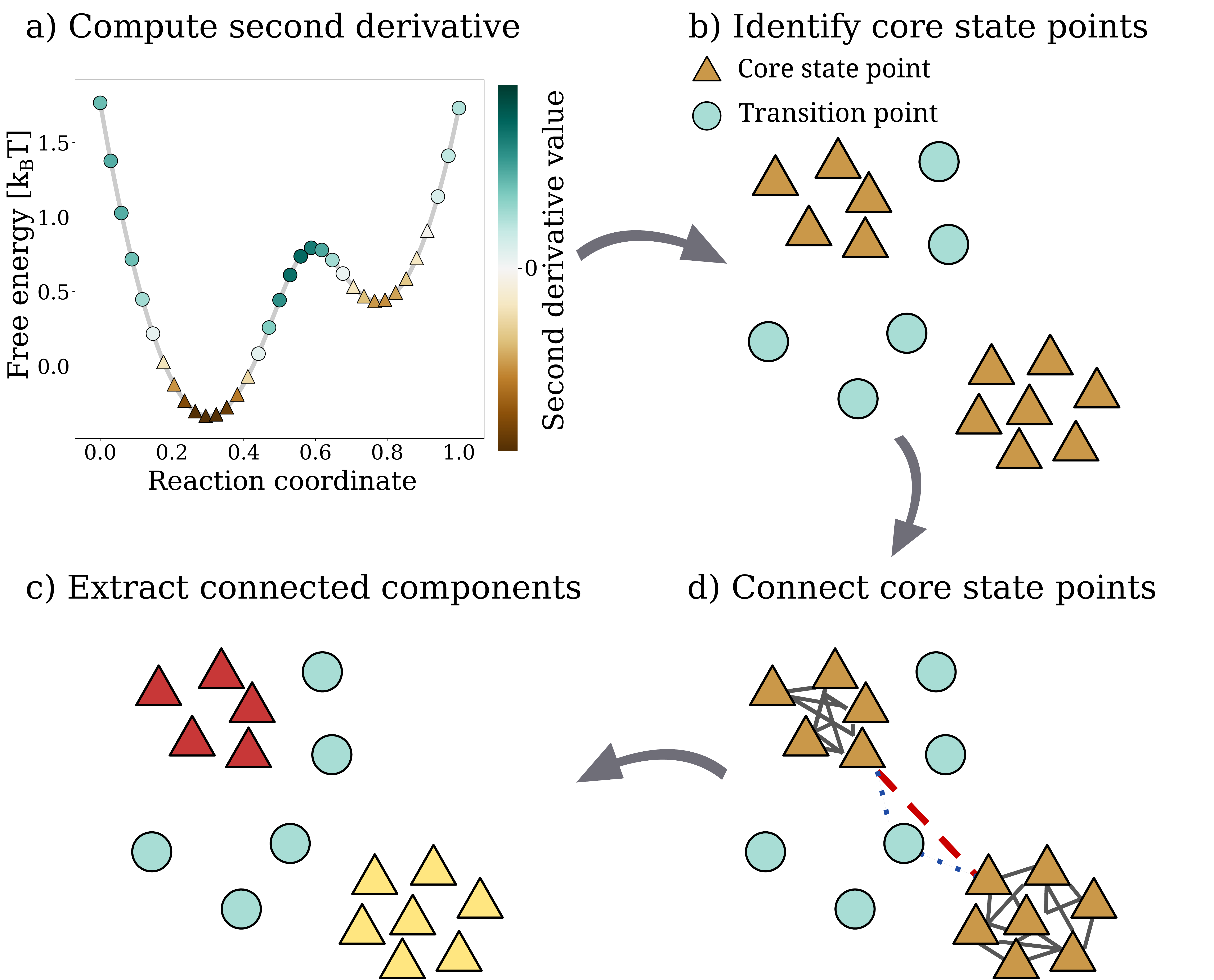}
\caption{An illustration of the InfleCS clustering method. First, the density Hessian is computed at all points. This is used to identify core state points. Connected graphs, or components, of core state points are then built using spatial proximity between core state points and transition state points. Finally, the connected components are extracted and cluster labels are assigned to the points in these components.}
\label{fig:second_derivative}
\end{figure} 

\subsection{Extracting core states from the density landscape}
Each point in a free energy landscape either belongs to a metastable core state or a transition state. To  analyze the conformational ensemble, we seek well-defined core states. Such core states are easily extracted from maxima of the estimated density landscape by exploiting its functional definition. The cutoff between core state and transition state is taken at the density inflection point.

Figure~\ref{fig:second_derivative} outlines the steps involved in the clustering method. In Figure~\ref{fig:second_derivative} a), a plot of a 1-dimensional 2-component Gaussian mixture free energy landscape is shown. The density second-order derivative values are displayed by colors. All points with negative density second derivative values are labeled metastable core states, while the rest are labeled transition points. Islands of core state points are isolated by transition points, Figure~\ref{fig:second_derivative} a,b). Two points within the same free energy minimum are connected by not allowing any transition point to be closer to both core state points. This creates connected components, that are extracted by assigning the same cluster label to all points within the same connected component, Figure~\ref{fig:second_derivative} c,d). 

To generalize the clustering to $N_\text{dims}$, we derive an expression for the density Hessian (matrix of second-order partial derivatives) with respect to the CVs. The partial derivative of a Gaussian mixture density, Eq.~\ref{eq:GMM_density}, with respect to the $d$th CV, $x_d$, is 
\begin{equation}
\frac{\partial }{\partial x_d} \rho_{\ve{a},\ve{\mu},\ve{\Sigma}} =  \sum\limits_{i=1}^{N_{\text{basis}}}a_i\;\mathcal{N}(x|\; \mu_i,\; \Sigma_i)\frac{\partial \hat{f}_i}{\partial x_d},
\end{equation}
where $\frac{\partial \hat{f}_i}{\partial x_d}$ is the $d$th element of the inner function gradient, $\nabla \hat{f_i} = -\Sigma_i^{-1}(x-\mu_i)$. 
From this we obtain an expression of the ($d$,$d'$)th element of the GMM density Hessian,
\begin{equation}
\frac{\partial^2 }{\partial x_d \partial x_{d'}} \rho_{\ve{a},\; \ve{\mu}, \; \ve{\Sigma}} = \sum\limits_{i=1}^{N_\text{basis}} a_i \; \mathcal{N}(x | \; \mu_i, \; \Sigma_i)\bigg( \frac{\partial^2 \hat{f}_i}{\partial x_d \partial x_{d'}} + \frac{\partial \hat{f}_i}{\partial x_d}\frac{\partial \hat{f}_i}{\partial x_{d'}} \bigg),
\end{equation}
where $ \frac{\partial^2 \hat{f}_i}{\partial x_d \partial x_{d'}} = -\Sigma^{-1}_{i,(d,d')}$. The Hessian reflects the curvature of the landscape, such that a point belongs to a metastable core state if the Hessian is negative definite. Since the Hessian is symmetric, this is the same as all its eigenvalues being negative. Thus, the shape of the density landscape is used to label each point as core state or transition state. 

The continuous definition of the density makes it possible to carry out the clustering on a grid instead of the sampled data, and subsequently use the grid clustering to assign labels to the sampled data. This is done by computing the Hessian of each grid cell center coordinate and mark each cell as core state or transition point, followed by the identification and extraction of connected components. Each sampled data point is then given the cluster label of the closest grid cell. This makes the core state extraction independent of the number of data points. The grid resolution mainly affects computational efficiency, and can be determined by the user, Figure S1. Here, we specify the resolution explicitly when a grid is used.

Transition points are left unassigned when identifying core states. For full clustering, the transition points are first sorted in order of descending density and one-by-one assigned to the closest cluster. The highest-density point of a cluster is taken as its cluster center. 

\subsection{Population of states}
To quantify the relative size of metastable states, we estimate the population of states, $\pi$. It reports on the probability of observing a configuration in any of the metastable states. The probability of the $k$th cluster is computed by integrating the density over its spanned volume, $V_k$,
\begin{equation}
\pi_k = \int\limits_{V_k} \rho_{\ve{a},\ve{\mu},\ve{\Sigma}}(x) \text{d} x = \int \limits_{X} I(x\in V_k) \rho_{\ve{a},\ve{\mu},\ve{\Sigma}}(x) \text{d} x.
\end{equation}
Here, $X$ is the full density domain and $I(x\in V_k)$ is an indicator function which is unity if $x\in V_k$, and zero otherwise. The integral is approximated with Monte Carlo integration with $10^5$ points sampled from the density landscape.


\section{Methods}

\subsection{Conventional and state-of-the-art clustering methods}
To evaluate performance and properties of InfleCS, its full clustering is compared to K-Means, agglomerative-Ward, Spectral clustering, hierarchical DBSCAN (HDBSCAN), density peaks advanced (DPA), robust density clustering (RDC), and canonical GMM. Since the number of clusters is usually not known in real-world datasets, we use simple heuristics that are based on the data to determine these. Clustering and heuristics for K-means, Agglomerative-Ward and GMM are obtained with scikit-learn~\cite{scikit-learn}. HDBSCAN is performed with the python package HDBSCAN~\cite{McInnes_hdbscan_2017}, while DPA and RDC are performed with tools from the original authors~\cite{derrico_automatic_2018,sittel_robust_2016}.

\subsubsection{K-means}
K-Means clustering is done by repeatedly assigning points to the cluster label corresponding to the nearest cluster center and updating the cluster center to the new cluster centroid. The silhouette score~\cite{rousseeuw_silhouettes_1987} is used to select the number of clusters. A high silhouette score indicates small within cluster distances and large distances to the closest cluster, and thus a good partitioning of spherical clusters. 

\subsubsection{Agglomerative-Ward}
Agglomerative-Ward (AW) clustering initially treats all data points as separate clusters. In each iteration, two clusters are merged to minimize within-cluster variance. This is similar to K-Means, but the cluster assignment is greedy while K-Means is optimized globally. Just as for K-Means, we determine the number of clusters with the silhouette score.

\subsubsection{Spectral clustering}
Spectral clustering makes use of local relationships by passing the data through a Gaussian kernel. Here, the Gaussian standard deviation is set to the maximum nearest neighbor distance to ensure that no point is disconnected. The processed data is used to create a random walk matrix from which the $K$ largest eigenvectors are identified. The row-normalized matrix with eigenvectors represents the embedding on a $K$-dimensional hypersphere. The embedded points are then clustered with K-means. 

The silhouette score is not easily applied to spectral clustering because the clustering is done in different spaces of spectral embeddings, which requires a non-trivial normalization~\cite{ponzoni_spectrus:_2015}. Instead, the largest eigengap is used to determine the number of clusters. 

\subsubsection{Hierarchical DBSCAN}
Hierarchical DBSCAN (HDBSCAN) extends the conventional DBSCAN clustering~\cite{ester_density_based_1996}, where core points are defined as points with at least $N_\text{neighbors}$ within a cutoff, $\varepsilon$. Connected components defining the clusters are built by connecting two core points if they belong to each other's $\varepsilon$-neighborhood. Because the fixed cutoff makes it difficult to identify clusters of different densities, HDBSCAN~\cite{campello_density-based_2013,McInnes_hdbscan_2017} instead hierarchically represents the DBSCAN partitions of varying $\varepsilon$ in a dendrogram. Stable clusters are then extracted through local cluster tree cuts~\cite{hartigan_consistency_1981}. 

To successfully extract clusters, a minimum allowed cluster size is required. It takes on the same value as $N_\text{neighbors}$. The default value of $N_\text{neighbors}=5$ is used on the toy model datasets.

\subsubsection{Density peaks advanced clustering}
The original density peaks~\cite{rodriguez_clustering_2014} is a recently developed method based on density estimation with local discrete basis functions. A decision graph is used to pick cluster centers with relatively high density and large distance to the closest point of higher density. The remaining points are then assigned to the closest cluster in order of decreasing density. This subjective picking of cluster centers, however, can be hampered by ambiguity, Figure~S2. Therefore, the density peaks advanced (DPA)~\cite{derrico_automatic_2018} clustering instead represents the data hierarchically and extracts clusters based on the data and significance of peaks. 

Whether or not a peak is significant is dictated by the parameter $Z$. A lower $Z$ value yields more sensitivity to density variations, which may result in identifying false clusters where the finite sampling leads to spurious fluctuations. A higher value of $Z$, on the other hand, decreases the sensitivity to density fluctuations, but may instead result in unidentified clusters. To pick $Z$, we plot the resulting number of clusters as a function of $Z$-values and choose $Z$ where the number of clusters plateaus, Figure~S3. 

In addition to the hierarchical representation, DPA uses a point adaptive $k$ nearest neighbor (PA$k$~\cite{rodriguez_computing_2018}) estimator to estimate the density and free energy along the data manifold. This requires a method to accurately estimate the intrinsic dimension of the dataset~\cite{granata_accurate_2016,facco_estimating_2017}. To carry out the full DPA pipeline, we used the TWO-NN method~\cite{facco_estimating_2017}. 


\subsubsection{Robust density clustering}
The robust density clustering~\cite{sittel_robust_2016} (RDC) was developed for clustering free energy intermediate states. First, the density is estimated by counting the number of points within a radius $R$ of each point. This yields a free energy estimate of each point. Starting with a low free energy cutoff, all points below the cutoff that are within a distance, $d_\text{lump}$, of each other are joined. The free energy cutoff is then iteratively increased by 0.1 $k_BT$ until all points are lumped together. This lumping procedure yields a separation of the clusters at free energy barriers. 

The lumping threshold, $d_\text{lump}$, is set to twice the mean of nearest neighbor distances~\cite{sittel_robust_2016}. Furthermore, guided by recent developments, we set $R=d_\text{lump}$~\cite{nagel_dynamical_2019}. Lastly, a minimum cluster size is required. To pick this parameter, we plot the resulting number of clusters as a function of the parameter, and choose a value where the number of clusters plateaus~\cite{nagel_dynamical_2019}, Figure~S4. 

\subsubsection{Canonical Gaussian mixture model clustering}
Canonical Gaussian mixture model clustering is done by fitting a Gaussian mixture density to the data, where each Gaussian component represents a cluster. A data point is assigned the cluster label corresponding to the component that contributes the most to its density. The number of components, and thus number of clusters, is chosen with BIC~\cite{mclaclan_number_2014, schwarz_BIC_1978}.

\subsection{Evaluating clustering on toy models}
Three two dimensional toy models  are used to compare and quantify performance of the different clustering methods. The first toy model is a dataset with seven Gaussian clusters. The clusters are non-equidistantly spaced and have different densities. The second dataset consists of three well-separated clusters, with one clearly non-Gaussian cluster. The third toy model attempts to mimic a real-world dataset with three poorly separated and nonlinear clusters with different densities and sizes. For each toy model, the minimum number of clusters (or Gaussian components) was set to 2, and the maximum to 15, for K-means, AW, spectral clustering, GMM and InfleCS.

Clustering quality can be assessed by computing the fraction of clustered points that originate from the same true class. This is the homogeneity score. A maximum homogeneity score is reached when the clustering is perfect, but also if points from one true class are divided into more than one cluster. A remedy is to instead report on the fraction of points from a true class that belongs to a single cluster, the completeness score. However, a perfect score is then obtained if points from different true classes are assigned to the same cluster. Since this is complementary to the homogeneity score, we use the average of the homogeneity and completeness scores, the so called V-measure~\cite{scikit-learn,rosenberg_vmeasure_2007}, to evaluate clustering quality. It gives a score between zero and one, where one indicates perfect clustering. To gather statistics, we repeat the sampling, clustering and V-measure evaluation 50 times for each toy model and clustering method. 

To further characterize InfleCS in terms of computational time and accuracy depending on the amount of added noise, number of data points, dimensionality and number of grids cells, we use simple clustering datasets (''blobs'', scikit-learn~\cite{scikit-learn}). We also used a nonconvex cluster dataset (''moons'', scikit-learn~\cite{scikit-learn}) to highlight InfleCS properties. 

\subsection{Molecular simulations of Ca\textsuperscript{2+}-bound Calmodulin}
We apply InfleCS to an ensemble of Ca\textsuperscript{2+}-bound Calmodulin (CaM) configurations~\cite{stevens_calmodulin_1983}, with minimum number of Gaussian components set to 10 and maximum to 25. CaM consists of 148 residues arranged in eight helices and three domains; the N-terminal and C-terminal lobes, as well as the flexible linker between them. The two lobes have two EF-hand motifs each, enabling CaM to bind four Ca\textsuperscript{2+} ions. The helices are named from A to H, from N- to C-terminus. In the Ca\textsuperscript{2+}-bound state, CaM commonly adopts a dumbbell shaped conformation with exposed hydrophobic clefts. The exposed hydrophobic clefts facilitate binding to, and thus regulating, a wide range of target proteins, including ion channels and kinases~\cite{kanehisa_kegg:_2017}.

We investigate the conformational dynamics of Ca\textsuperscript{2+}-bound CaM with 460 ns already published~\cite{westerlund_effect_2018} replica exchange solute tempering (REST) simulations~\cite{wang_replica_2011,bussi_hamiltonian_2014}. To analyze CaM, we project the protein heavy atom coordinates onto two collective variables. The first CV, difference in distribution of reciprocal interatomic distances (DRID)~\cite{zhou_distribution_2012}, reflects on the global conformational change relative the initial crystal structure~\cite{babu_structure_1988}. In short, the distribution of inverse distances between $C_\alpha$ atoms are used to extract three features for each residue; the mean, the square root of the second central moment, and the cubic root of the third central moment. Thus, each frame $j$ is described by a feature matrix, $\ve{v}_j\in \mathbb{R}^{3\times N_{\text{Residues}}}$. The difference to the initial structure with features $\ve{v}_0$ is computed as the average residue feature distance
\begin{equation}
\text{DRID}_j = \frac{1}{3N_\text{Residues}} \sum\limits_{n=1}^{N_\text{Residues}} \| \ve{v}_j^n-\ve{v}_0^n \|.
\end{equation}

The second CV, backbone dihedral angle correlation (BDAC)~\cite{westerlund_inference_2018}, reports on secondary structure changes in the linker relative the initial structure,
\begin{equation}
\text{BDAC}_j = \frac{1}{4N'_\text{Residues}}\sum\limits_{n=1}^{N'_{\text{Residues}}} (2 + \cos (\varphi_j^n - \varphi_0^n) + \cos (\psi_j^n - \psi_0^n) ).
\end{equation}
Here, $\varphi_j^n$ and $\psi_j^n$ are the backbone dihedral angles of linker residue $n$ in frame $j$. MDtraj~\cite{mcgibbon_mdtraj:_2015} was used to compute DRID feature vectors and backbone dihedral angles.

\section{Results and discussion}

\subsection{Toy models demonstrate properties of InfleCS}
\begin{figure}
    \centering
    \includegraphics[width=0.7\paperwidth]{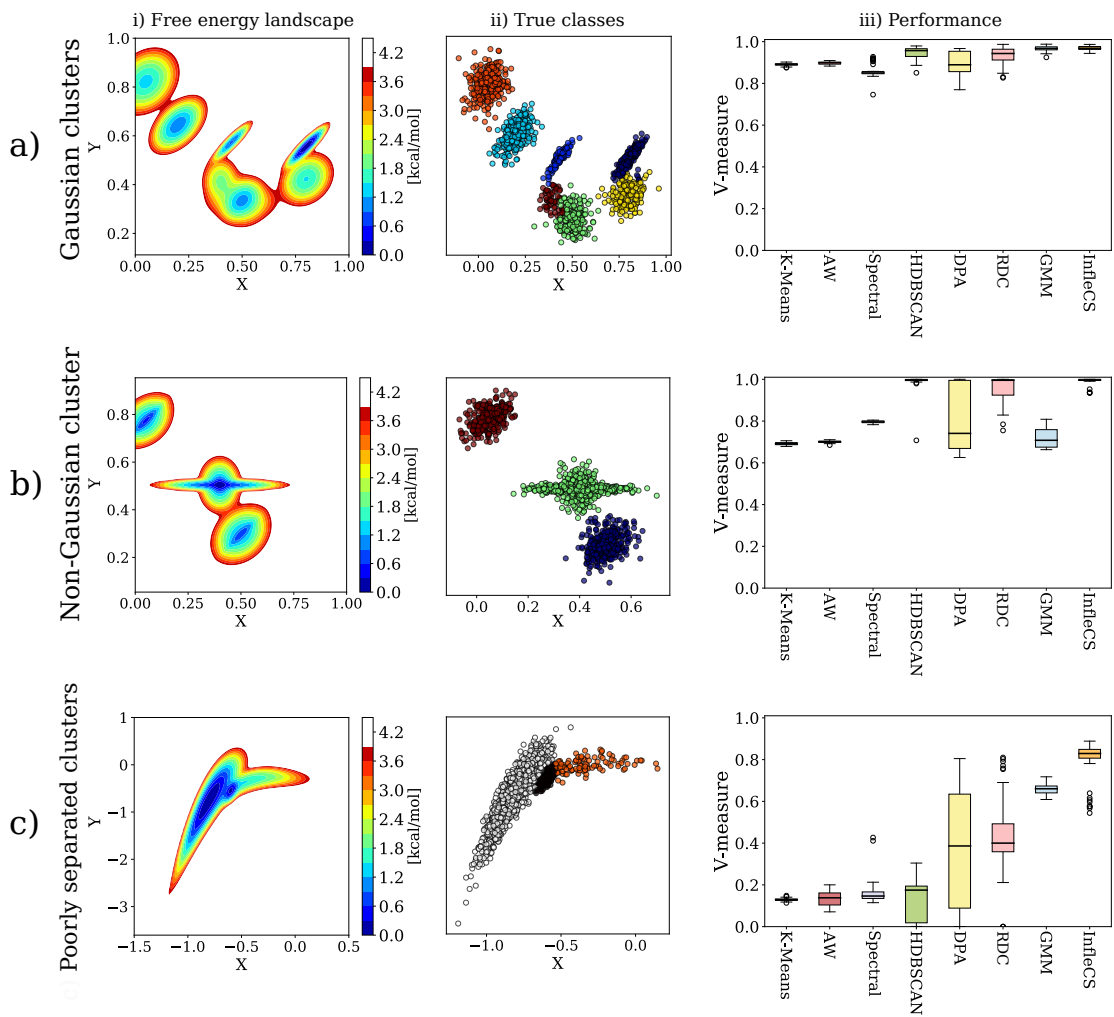}
    \caption{The performance of all clustering methods (K-Means, Agglomerative-Ward (AW), spectral clustering, HDBSCAN, density peaks advanced (DPA), robust density clustering (RDC), Gaussian mixture model (GMM) and InfleCS) on three toy model datasets with a) Gaussian clusters, b) Non-Gaussian clusters and c) hierarchical and poorly separated clusters. For each toy model, we show i) the true free energy landscape of the toy model, ii) an example of sampled data and its true clustering and iii) box plots showing the V-measure scores for the clustering methods. Each box is constructed with 50 sampled datasets. The horizontal line of each box shows the median, while the box covers the region between the lower quartile $Q1$ (25th percentile) and the upper quartile $Q3$ (75th percentile). The whiskers mark data obtained within $Q1-1.5IQR$ and  $Q3+1.5IQR$, where $IQR=Q3-Q1$ is the interquartile range. Outliers are shown as dots.}
    \label{fig:toy_models}
\end{figure}

The first toy model consists of Gaussian clusters generated from the free energy landscape in Figure~\ref{fig:toy_models}~a.i). An example of sampled data with true cluster labels is displayed in Figure~\ref{fig:toy_models}~a.ii), and performances of the different clustering methods on this toy model are shown in Figure~\ref{fig:toy_models}~a.iii). Because the clusters are Gaussian and have similar spatial size, most methods perform well. GMM and InfleCS specifically provide close to perfect clustering. 

To investigate the impact of cluster shapes on clustering performance, a second toy model with one non-Gaussian cluster is introduced, Figure~\ref{fig:toy_models}~b.i-ii). This toy model highlights how GMM, K-means and agglomerative-Ward fail because their intrinsic assumptions are not met, Figure~\ref{fig:toy_models}~b.iii). DPA clustering is highly variable, indicating that the chosen peak significance value is not optimal across sampled datasets. Interestingly, spectral clustering has similar performance as on the first toy model, and the other methods, especially HDBSCAN and InfleCS, provide high performance clustering. The assumption of a density landscape that can be accurately described by a linear combination of Gaussian components is the core of GMM density estimation and, consequently, InfleCS clustering. Even so, most densities can be modeled by a GMM given enough sampled points and Gaussian components. Moreover, because InfleCS extracts clusters with graphs based on local relationships between data points, it is possible to identify non-Gaussian and, to some extent, even uniform and highly nonlinear clusters, as exemplified with the moons dataset, Figure~S5.


Most free energy landscapes rendering conformational ensembles of biomolecules, however, do not contain uniform or highly nonlinear states. The problem of identifying metastable states is then instead related to the hierarchical nature of the landscapes, as well as the noisy state definitions that occur when a high dimensional dataset is projected onto much fewer dimensions. The third toy model mimics such a dataset, where the originally high dimensional data is projected onto a low dimensional space with poorly separated clusters. By mere inspection of the scattered data, Figure~\ref{fig:toy_models}~c.ii), it is difficult to identify the clusters. The free energy landscape, however, clearly depicts three states, Figure~\ref{fig:toy_models}~c.i). Because the clusters are poorly separated and of different spatial sizes, the geometric (not density based) clustering methods completely fail, Figure~\ref{fig:toy_models}~c.iii). Furthermore, HDBSCAN, DPA and RDC have low clustering qualities due to the spatially small-sized clusters of relatively low density. Despite the poorly separated states with different density and sizes, InfleCS maintains high quality clustering, Figure~\ref{fig:toy_models}~c.iii). Thus, it is the only method in this set that successfully clusters all toy model datasets. 

In addition to projection issues, data sampled by molecular dynamics simulations may contain spurious noise. Uniformly distributed noise is relatively well tolerated by the InfleCS clustering, Figure~S6. Another important aspect to consider is the dimensionality of the data: in some cases, more than 5 dimensions are needed to describe the conformational ensemble of a protein~\cite{facco_estimating_2017,sittel_perspective_2018}. Due to the curse of dimensionality, datasets of larger dimensionality are required to contain more data points to allow for density estimation. Therefore, a dataset with larger dimensionality notably affects the computational time, which increases with both data dimensionality and the number of data points, Figure~S7~b) and~S8. Still, because GMMs use continuous basis functions and estimate densities globally, the estimates in sparse regions tend to be more accurate than those obtained from methods with discrete basis functions~\cite{wasserman_nonparametric_2004,westerlund_inference_2018}. Because InfleCS relies on such density estimation, it is possible to perform accurate clustering in higher dimensional space when enough data is available, Figure~S7~a).

\subsection{Application to Ca\textsuperscript{2+}-bound Calmodulin ensemble}
\begin{figure}
\centering
\includegraphics[width=0.7\paperwidth]{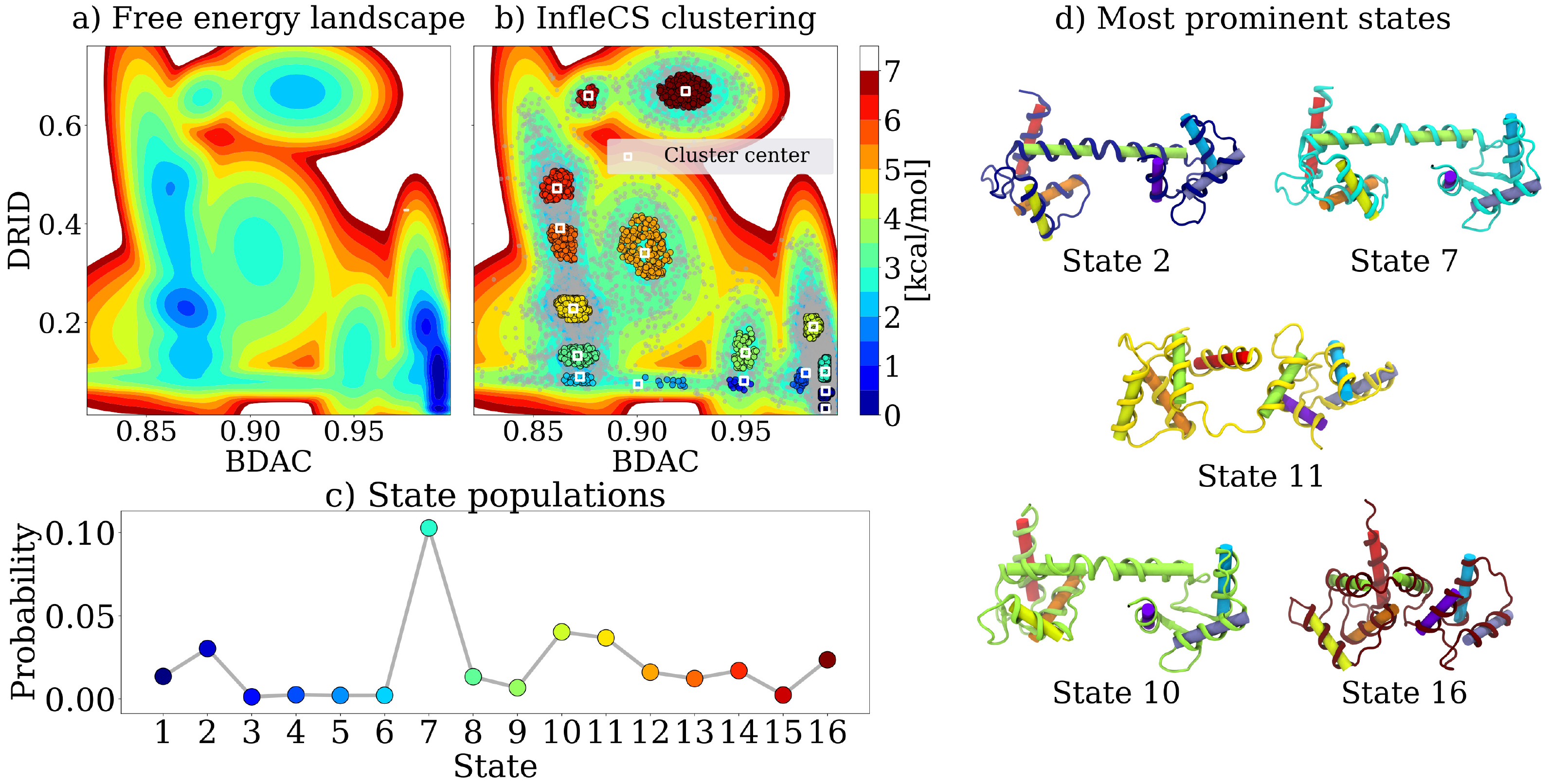}
\caption{a) The estimated free energy landscape of the CaM conformational ensemble along the distribution of reciprocal interatomic distances (DRID) and the linker backbone dihedral angle correlation (BDAC). The final density landscape consists of 16 Gaussian components. The estimated parameters of the 16-component GMM on this dataset are listed under Section~S2. b) The identified core states based on the estimated density on a $100\times 100$ grid, colored by cluster labels. Transition points are shown as gray dots. c) The state populations. d) Structures of each cluster center. The ribbons are colored according to the cluster the structure belongs to, while the cylinders, helix A to H, are colored according to a rainbow. The structures are visualized with VMD~\cite{humphrey_VMD_1996}.}
\label{fig:CaM_InfleCS}
\end{figure}

The estimated free energy landscape of CaM along the two CVs and the corresponding InfleCS clustered data are shown in Figure~\ref{fig:CaM_InfleCS} a-b). As expected, InfleCS correctly identifies the metastable states within the estimated free energy landscape. The complex nature of the CaM dataset and the stochasticity of GMM density estimation result in a relatively plateaued $I_{BIC}$ profile, Figure~S9~a). Nonetheless, by repeating the clustering 50 times, we remark that InfleCS maintains robust clustering, Figure~S9~b). Note that the clustering is carried out on a grid, the resolution of which has a negligible impact on the clustering, Figure~S1.

The core state probabilities, Figure~\ref{fig:CaM_InfleCS} c), show that the seventh state is the most common state in this dataset. Representative structures of the most populated states are shown in Figure~\ref{fig:CaM_InfleCS} d). The most common structure is in a canonical dumbbell conformation, but we also identify two different compact conformations among these well-populated states, namely state 11 and 16. Compact states similar to state 16 have been identified in previous MD datasets~\cite{aykut_designing_2013,fiorin_unwinding_2005,shepherd_molecular_2004,wriggers_structure_1998} as well as in an experimental structure~\cite{fallon_closed_2003}.

\begin{figure}
\centering
\includegraphics[width=0.5\paperwidth]{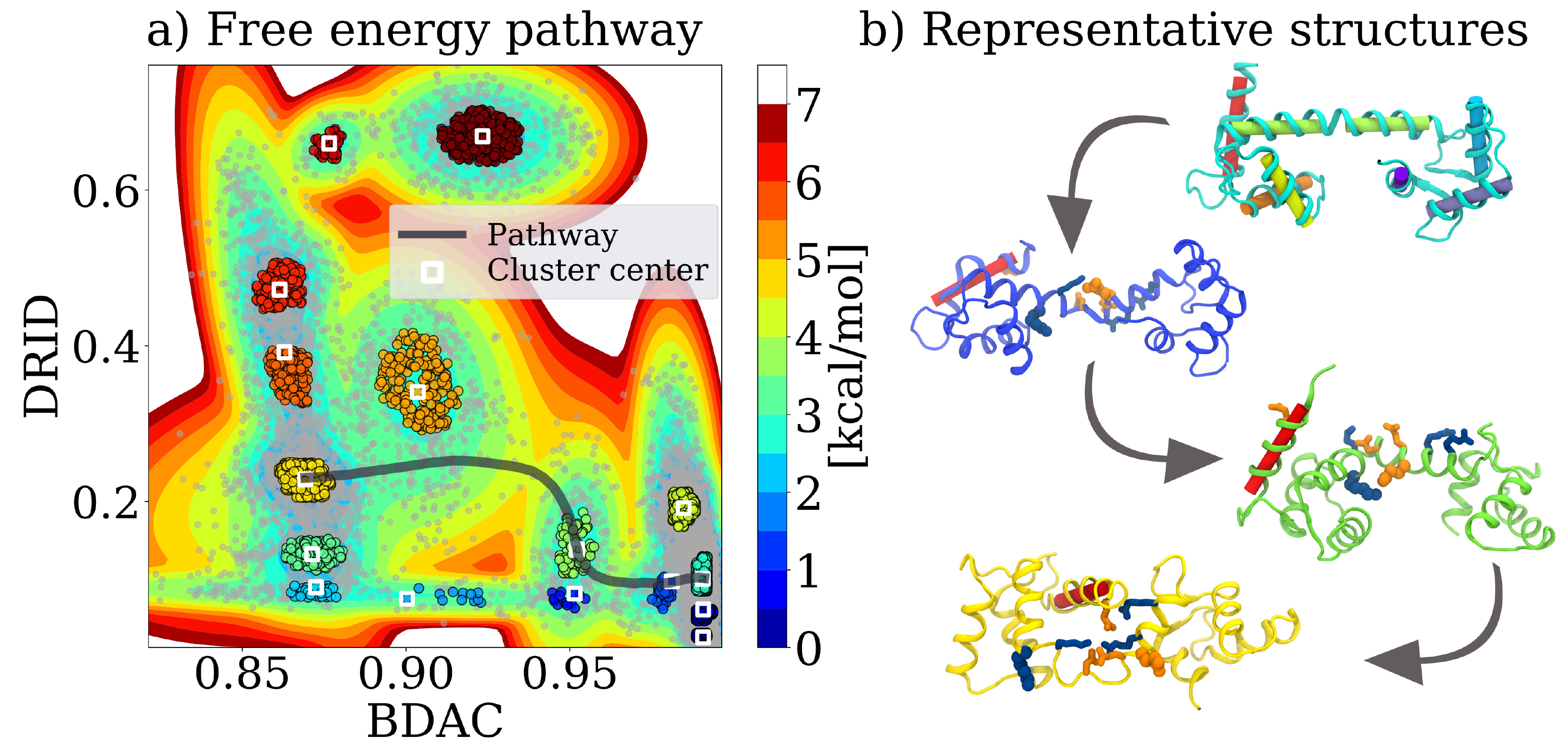}
\caption{a) A free energy pathway between the most common state 7 to the compact state 11. The path goes through state 4 and 9. b) Representative structures of the involved states. The ribbons are colored according to the cluster the structure belongs to, while the cylinder, helix A, is colored red. The positively (negatively) charged residues that participate in salt-bridge formation and breaking are shown with blue (orange) sticks (GLU11, LYS77, ASP80, GLU83, ARG86, ARG90). LYS75 and GLU84 are shown with blue and orange spheres, respectively. The structures are visualized with VMD~\cite{humphrey_VMD_1996}.}
\label{fig:CaM_pathway}
\end{figure}

With the estimated free energy landscape available, we can infer possible pathways between states and thus understand how the states are connected. For example, the eleventh state is a relatively highly populated compact state with helix A collapsed onto the linker. By mere inspection of the representative structures, it is not obvious how CaM transitions to this state from the canonical conformation (state 7). Through visual inspection of the free energy landscape, we characterize one possible pathway between the canonical and compact state that goes through state 4 and 9, Figure~\ref{fig:CaM_pathway}~a). The representative structures corresponding to cluster centers, Figure~\ref{fig:CaM_pathway} b), indicate that the transition along this pathway likely occurs through a twisting motion of the two lobes around the linker and breaking of the linker helix. 

Early MD simulations and experimental studies of CaM suggest that destabilization of the linker helix is driven by electrostatic interactions~\cite{fiorin_unwinding_2005,spoel_bending_1996, shepherd_molecular_2004,torok_effects_1992,kataoka_linker_1996}. We therefore study salt bridges in the three core state ensembles along the pathway. Going from the canonical to the intermediate state, the linker helix breaks which enables formation of a stabilizing salt bridge between LYS75 and GLU84, Figure~\ref{fig:CaM_pathway}~b) and~S10 a-c). To transition to the compact state, GLU11 is likely  recruited by LYS77, which breaks the LYS75-GLU84 salt bridge (cluster 9) in favor of a new charged cluster with salt bridges between LYS77-GLU83, ASP80-ARG86, GLU11-ARG86 and GLU11-ARG90, Figure~\ref{fig:CaM_pathway} b) and~S10~c-d). LYS75 is a common target-protein binding residue~\cite{villarroel_ever_2014}. Thus, the salt bridge formations in this compact state that expose LYS75 to solvent may promote initiation of long-ranged interactions with target proteins.

\section{Conclusions}
We presented InfleCS, a clustering method that uses the shape of an estimated Gaussian mixture density to identify metastable core states. The method was shown to consistently outperform other common clustering methods on three toy models with different properties. 

The advantages with InfleCS for free energy landscape clustering are five-fold. First, clusters are identified at density peaks, which guarantees that clusters are metastable states. Second, core state boundaries identified by density second derivatives result in well-defined states. Third, clusters are constructed by building graphs, thus making less assumptions about cluster shapes. Fourth, because the clustering method naturally involves density and free energy landscape estimation, it is possible to derive pathways between states and thus understand fundamental mechanisms. Finally, the number of clusters is naturally determined by the number of density peaks in the landscape and therefore requires no a priori system knowledge. 

By applying InfleCS to a conformational ensemble of Ca\textsuperscript{2+}-bound CaM, we identified a possible pathway from the canonical to a compact state through a twisting motion of the two lobes followed by salt bridge breaking and formation. This pathway highlights electrostatically driven structural rearrangements that may allow CaM to bind to a wide range of target proteins.

\begin{acknowledgement}
The simulations were performed on resources provided by the Swedish National Infrastructure for Computing (SNIC) at PDC Centre for High Performance Computing (PDC-HPC). This work was supported by grants from the Science for Life Laboratory.
\end{acknowledgement}

\suppinfo
The code and tutorial for estimating free energy landscapes and cluster with InfleCS is available free of charge at http://www.github.com/delemottelab/InfleCS-free-energy-clustering-tutorial. 

\newpage

\bibliography{zotero}


\end{document}